\documentclass[fleqn]{article}

\usepackage{fancybox}
\usepackage{amssymb}
\usepackage{amsfonts}
\usepackage{amsmath}
\usepackage[amsmath,thmmarks]{ntheorem}
\usepackage{scalefnt}
\usepackage{url}
\usepackage{verbatim}
\usepackage[dvipsnames]{xcolor}
\usepackage{nccmath}
\usepackage{eurosym}
\usepackage{paralist}
\usepackage{hyperref}
\usepackage{authblk}
\usepackage[top=1in, bottom=1.25in, left=1.25in, right=1.25in]{geometry}

\usepackage{xspace}
\usepackage[latin1]{inputenc}

\usepackage[ruled,vlined,linesnumbered]{algorithm2e}

\usepackage{tikz}
\usepackage{tkz-graph}
\usetikzlibrary{arrows,automata,positioning}
\usetikzlibrary{shapes}

\newtheorem{theorem}{Theorem}

{\theorembodyfont{\upshape}
  \theoremsymbol{\ensuremath{\Diamond}}

  \theoremsymbol{}
  \newtheorem{example}[theorem]{Example}
  \theoremstyle{nonumberplain}
  \theoremsymbol{}
\theoremseparator{.}
  \newtheorem{remark}{Remark}
}
\theoremheaderfont{\sc}\theorembodyfont{\upshape}
\theoremstyle{nonumberplain}
\theoremseparator{.}
\theoremsymbol{\rule{1ex}{1ex}}

\newcommand{\isdef}{\stackrel{\mbox{\tiny def}}{=}}

\newcommand{\set}[1]{\left\{#1\right\}}

\newcommand{\setof}[2]{\left\{#1\,\middle|\:#2\right\}}

\newcommand{\dN}{\mathbb{N}}

\newcommand{\dR}{\mathbb{R}}
\newcommand{\dB}{\mathbb{B}}

\newcommand{\calA}{\mathcal{A}}
\newcommand{\calM}{\mathcal{M}}
\newcommand{\calN}{\mathcal{N}}
\newcommand{\keyw}[1]{\mathbf{#1}}
\newcommand{\hide}[1]{}

\begin{document}
	\title{Formalizing the Cox-Ross-Rubinstein pricing of European derivatives in Isabelle/HOL}

\author[1,2]{Mnacho Echenim\thanks{\texttt{Mnacho.Echenim@univ-grenoble-alpes.fr}}}
\author[1,3]{Herv\'e Guiol\thanks{\texttt{Herve.Guiol@univ-grenoble-alpes.fr}}}
\author[1,2]{Nicolas Peltier\thanks{\texttt{Nicolas.Peltier@univ-grenoble-alpes.fr}}}
\affil[1]{Univ. Grenoble Alpes, F-38000 Grenoble France}
\affil[2]{CNRS, LIG}
\affil[3]{CNRS, TIMC}

\date{July 2018}

\maketitle

\begin{abstract}
We formalize in the proof assistant Isabelle essential basic notions and results in financial mathematics.
We provide generic formal definitions of concepts such as markets, portfolios, derivative products, arbitrages or fair prices, and we show that, under the usual no-arbitrage condition, the existence of a
replicating portfolio for a derivative implies that the latter admits a unique fair price.
Then, we provide a formalization of the Cox-Rubinstein model and we show that the market is complete in this model, 
i.e., that every derivative product admits a replicating portfolio. This entails that in this model, every derivative product admits a unique fair price.
\end{abstract}

\section{Introduction}

%

The basic securities that are traded on financial markets --such as shares on the equity market or bonds on the fixed-income market-- have a price that is submitted to the law of supply and demand, and depends on the needs of financial actors. Things are not that simple for all securities traded on financial markets, and in particular, determining the price of so-called \emph{derivative products} can be a far from trivial task. A derivative product is a security the value of which depends on that of one or several underlying securities; a typical example is a vanilla call on a share, which gives its holder the right, but not the obligation, to buy the share on a predetermined date at a predetermined price. Obviously, the price of a derivative product should depend on that of its underlyings, but what exactly is this dependency? In fact, is there even a unique price for any derivative? An intuitive answer to the second question is that the price of a derivative should be unique: if this were not the case, an investor could buy the derivative at the lower price and simultaneously sell it at the higher price, making a profit without investing any money or taking any risks. The investor would have exploited what is called an \emph{arbitrage opportunity} and although such opportunities do exist on financial markets, they are exploited by financial actors called arbitragists and tend to disappear quickly. This is the reason why many results in quantitative finance rely on a no-arbitrage hypothesis. Such a hypothesis also permits to provide a more precise definition of what a price for a derivative should be: this should be any value that is neither so high as to induce an arbitrage opportunity for the seller of the derivative, nor so low as to induce an arbitrage opportunity for the buyer. Any price satisfying these conditions is called a fair price for the derivative.

One of the most important results in financial mathematics was the proof by Black, Scholes and Merton \cite{BlackScholes,Merton} that, in the so-called Black-Scholes model of an equity market, every derivative admits a unique price, along with a formula permitting to compute this price, either numerically or explicitly. Along with the no-arbitrage hypothesis, the authors assume that 
\begin{inparaenum}[(1)]
	\item the market is \emph{frictionless}, meaning that securities can be bought or sold with no transaction costs, and 
	\item investors can buy and sell any amount of the securities, meaning that the quantity of a security withheld in a portfolio can be any real number, even a negative one if the security has been sold short (i.e., sold by an investor not owning the asset).
\end{inparaenum}
Since then, there have been a wide variety of mathematical models devised for the pricing of derivative products, adapting the hypotheses of the Black-Scholes-Merton model or modeling other markets, such as the foreign-exchange or commodities markets. 

A discrete-time model of an equity market was introduced in 1979 by Cox, Ross and Rubinstein \cite{CRR}. This model is based on hypotheses similar to those of the Black-Scholes-Merton model, in which time is continuous, and can actually be viewed as a discrete-time approximation of this model. The complexity of evaluating the price of a derivative in this model implies that it is not frequently used for the pricing of simpler derivatives. But several financial institutions still rely on this model for the pricing of more complex derivative, such \emph{american options}, which can be exercised by their buyer at \emph{any} time until a given maturity.

In this paper, we present a formalization in Isabelle/HOL \cite{Nipkow:2002:IPA:1791547} of 
\begin{inparaenum} [(1)]
	\item fair prices {for derivative products} on equity financial markets, 
	\item the proof of their uniqueness when a replicating portfolio exists in a fair market, and
	\item  an algorithm to compute fair prices under a risk-neutral probability space.
\end{inparaenum}
We also formalize the Cox-Ross-Rubinstein model and prove that in this model, every derivative admits a replicating portfolio, i.e., a portfolio with a value at maturity identical to the payoff of the derivative. The work presented here strictly subsumes the formalization carried out in \cite{EPCRR}, which was mainly devoted to the proof that in the model of a market defined by Cox, Ross and Rubinstein \cite{CRR}, every derivative product admits a replicating portfolio. The results presented in this paper can be found in many financial mathematics textbooks, with one main difference: in general, the results are presented by considering an arbitrary derivative with a given maturity $T$, and taking the finite probability space with outcomes consisting of all sequences of $T$ coin tosses. Here we formalize a setting in which \emph{any} derivative can be priced, and use Isabelle's codatatypes \cite{Blanchette13} to consider non-denumerable probability space with outcomes consisting of all infinite streams of coin tosses.

\paragraph{Related work.} Many results related to financial mathematics have already been formalized in Isabelle. Large parts of Probability theory have been formalized, building up on \cite{hoelzl2012thesis}; and results and concepts frequently used in financial mathematics such as Markov processes or the Central Limit Theorem are also available in Isabelle \cite{HolzlJAR,CentralLimit}. To the best of our knowledge, other than \cite{EPCRR}, there has been no formalization of financial mathematics.

\paragraph{Organization.} This paper is organized as follows. Section \ref{sect:preliminaries} contains basic financial notions as well as a summary of the notions from Probality theory that will be used throughout the paper and are already formalized in Isabelle. In Section \ref{sect:equity_market}, we define equity markets in discrete time, introducing the notions of portfolios and their values, as well as trading strategies which represent the only reasonable portfolios that can be constructed. Arbitrage opportunities are introduced in Section \ref{sect:fair}, they permit to define the notion of af air price for a derivative, and we show that if a derivative admits a portfolio whose value at maturity is identical to the derivative payoff, then the fair price for this derivative is unique. Section \ref{sect:riskneutr} is devoted to the definition of risk-neutral probability spaces, which are based on the existence of martingales, and permit to represent the fair price of a derivative as an expectation. In Section \ref{sect:CRR}, these results are applied to the Cox-Ross-Rubinstein model, and an explicit formula for computing the fair price of any derivative is provided. Section \ref{sect:comp_ex} contains a detailed illustrative example, showing how in the Cox-Ross-Rubinstein model, a replicating portfolio is computed, and the fair price of a derivative is obtained. The theory files described in this paper are available on the \emph{Archive of Formal Proofs}, at \url{https://www.isa-afp.org/entries/DiscretePricing.html}.

\section{Preliminary notions}

\label{sect:preliminaries}

\subsection{Some notions in finance}

We begin by briefly reviewing some basic standard definitions about equity markets. This treatment is mainly based on Shreve \cite{Shreve}, Vol. 1.
An equity market consists of a set of \emph{assets} or \emph{securities} that can be traded at prices that evolve with time.
An actor trading on different assets will own a \emph{portfolio} containing different quantities of the traded assets. These quantities are real numbers that can be positive if the corresponding asset was bought, or negative if the asset was the object of a short sale. A portfolio can be \emph{static} if its composition is fixed once and for all, and  \emph{dynamic} if its composition can evolve over time. Clearly, almost all portfolios on markets are dynamic ones. Among the dynamic portfolios, those of a particular interest are the \emph{trading strategies}; these are the dynamic portfolios for which the composition at time $t$ is a random variable that only depends on the available information up to time $t$; trading strategies are thus meant to represent portfolios for which no insider trading can occur. 
A portfolio in which cash is only invested at inception, after what all future trades are financed by buying or selling assets in the portfolio is a \emph{self-financing portfolio}. An arbitrage represents a ``free lunch'': it is defined as a self-financing trading strategy with a 0 initial investment that offers a risk-free possibility of making a profit. A market is \emph{viable} if it offers no arbitrage opportunities.

Some of the securities that can be traded are basic securities, such as bonds, which are generally assumed to be risk-free assets, or stocks, which are risky assets. Others are \emph{derivative} securities, with payoffs (the amount of cash that should be exchanged at exercise time)
 that depend on the evolution and values of underlying securities.
On the equity market, these derivative securities often have an expiry date, or maturity, after which they are no longer valid. 
 An option, for instance, is a derivative that can be viewed as an insurance: when it is exercised, it gives the beholder the right---but not the obligation---to trade an instrument at a given price. In this paper we will focus on {\em European} options, which can only be exercised at the maturity, see, e.g., \cite{Hull}.
The best-known options are the call and the put options. A call (resp. put) option gives its beholder the right, at time $T$, to buy (resp. sell) the underlying security at the strike price $K$, thus guaranteeing that there is a cap (resp. floor) on the price that will be payed at a future time for the security. In practice, when at time $T$ the price of the underlying security, denoted by $S_T$, is greater than the strike price $K$, the buyer of a call receives $S_T - K$ from the seller of the option, and buys the instrument on the market for $S_T$, in effect only spending $K$ to obtain the instrument. When $S_T < K$, the seller of the call does not deliver any cash, as the buyer will directly buy the instrument on the market for a value that is less than $K$. Thus, a call option is a derivative that, at maturity $T$, delivers a payoff 
of $(S_T -K)^+ \isdef \max (0, S_T-K)$. In a similar way, a put option delivers at time $T$ a  payoff of $(K-S_T)^+$.

Once a derivative is sold, the seller is meant to invest the cash by creating a trading strategy, in order to be able to pay the required amount of money when the derivative is exercised.
A natural question is the following: how much should a buyer be expected to pay for a given derivative? Ideally, this price should not be so low that the buyer could make a riskless profit, and it should not be so high that the seller could make a riskless profit. As we will see, the answer to this question is quite straightforward when the seller is capable of creating a trading strategy that generates at exercise time exactly the payoff of the derivative, i.e., of creating a replicating portfolio. In this case, the fair price for the derivative is the investment needed to initiate the trading strategy. A market is \emph{complete} if every derivative admits a replicating portfolio; in a complete market, every derivative admits a fair price.

The construction of replicating portfolios is clearly not straightforward, and it may not be guaranteed that such portfolios actually exist.
An answer to the existence of replicating portfolios for European options was given by Fischer Black and Myron Scholes, and by Robert Merton in \cite{BlackScholes,Merton}, in the so-called Black-Scholes-Merton model.
They consider a risky asset, the stock, that pays no dividends and whose evolution is described by a \emph{geometric Brownian motion} (see, e.g., \cite{Hull}), and showed that, under some simple market hypotheses such as identical bidding and asking prices for the stock and the absence of arbitrage opportunities, a European option over a single stock can be replicated with a portfolio consisting of the stock and a cash account. Their proof is based on the construction of a dynamic portfolio, the composition of which changes continuously (it is called a \emph{delta-neutral portfolio}), which is guaranteed to replicate the option  under consideration. Along with the construction of replicating portfolios, the authors provide a formula that permits to compute the fair price of any European option.

The Cox-Ross-Rubinstein model \cite{CRR} that we consider in Section \ref{sect:CRR} of this paper can be viewed as an approximation of the Black-Scholes-Merton model to the case where time is no longer continuous but discrete; i.e., to the case where securities are only traded at discrete times $1,2,\ldots, n,\ldots$ In this setting, the evolution of the stock price is described by a geometric random walk, which can be viewed as a discrete version of the geometric Brownian 
motion: if the stock has a price $s$ at time $n$, then at time $n+1$, this price is either $u.s$ (upward movement) or $d.s$ (downward movement). The probability of the price going up is always $0 < p < 1$, and the probability of it going down is $1-p$.
The authors show that under these conditions, the market is complete: every derivative admits a replicating portfolio.

\subsection{Probability theory in Isabelle: existing notions}

A large part of the formalization of measure and probability theory in Isabelle was carried out by Hölzl \cite{hoelzl2012thesis} and is now included in Isabelle's distribution.  We briefly recap some of the notions that will be used throughout the paper and the way they are formalized in Isabelle. We assume the reader has knowledge of fundamental concepts of measure and probability theory; any missing notions can be found in Durrett \cite{Durrett} for example. 
For the sake of readability, in what follows, a term $F\, t$ will sometimes be written $F_t$.

\newcommand{\sigsets}{\texttt{sigma-sets}}

Probability spaces are particular \emph{measure spaces}.  A measure space over a set $\Omega$ consists of a function $\mu$ that associates a nonnegative number {or $+\infty$} to  some subsets of $\Omega$. {The subsets of $\Omega$ that can be measured are closed under  complement and countable unions and make up a \emph{$\sigma$-algebra}.} 
The $\sigma$-algebra \emph{generated by a set} ${C} \subseteq 2^\Omega$ is the smallest $\sigma$-algebra containing ${C}$; it is denoted in Isabelle by $\sigsets\ \Omega\ {{C}}$. 

The functions $\mu$ that measure the elements of a $\sigma$-algebra are positive and \emph{sigma additive}\footnote{This property is also called \emph{countable additivity} in the literature.}: if $\calA \subseteq 2^\Omega$ is a $\sigma$-algebra and the sequence $(A_i)_{i\in \dN}$ consists of pairwise disjoint elements in $\calA$, then $\mu(\bigcup_{i\in\dN} A_i) = \sum_{i\in \dN} \mu(A_i)$. In Isabelle, measure spaces are defined as follows (where $\overline{\dR}$ denotes $\dR \cup \{ -\infty,+\infty \}$ and $\dB = \{ \bot, \top \}$):
\newcommand{\measurespace}{\texttt{measure-space}}
\newcommand{\sigmaalgebra}{\texttt{$\sigma$-algebra}}
\newcommand{\ispositive}{\texttt{positive}}
\newcommand{\countablyadditive}{\texttt{countably-additive}}
\newcommand{\isaset}{\texttt{set}}
\newcommand{\measure}{\texttt{measure}}
\newcommand{\isaspace}{\texttt{space}}
\newcommand{\sets}{\texttt{sets}}
\newcommand{\emeasure}{\texttt{emeasure}}
\newcommand{\measurable}{\texttt{measurable}}
\newcommand{\borelmeasurable}{\texttt{borel-measurable}}

\[\begin{array}{lcl}
\measurespace & :: & \alpha\,\isaset \rightarrow \alpha\,\isaset\,\isaset\rightarrow \left(\alpha\,\isaset \rightarrow \overline{\dR}\right) \rightarrow \dB\\
\measurespace\ \Omega\ \calA\ \mu & \Leftrightarrow&\\
 {\quad \sigmaalgebra\ \Omega\ \calA\ \wedge\ \ispositive\ \calA\ \mu\  \wedge\ \countablyadditive\ \calA\ \mu}\span \span
\end{array}\]
A measure type is  defined by fixing the measure of non-measurable sets to $0$:
\[\begin{array}{l}
\keyw{typedef}\ \alpha\ \measure  =
{\setof{(\Omega, \calA, \mu)}{(\forall A\notin \calA.\, \mu A = 0) \wedge \measurespace\ \Omega\, \calA\, \mu}}
\end{array}\]
If $\calM$ is an element of type $\alpha$ {\measure}, then the corresponding space, $\sigma$-algebra and measure are respectively denoted by $\Omega_\calM$, $\calA_\calM$ and $\mu_\calM$. 

 The definition of a measure type may seem surprising, especially to mathematicians, because setting the measure of a set not in $\calA$ to $0$ can be counter-intuitive: there is for example no relationship between elements with a measure 0 and negligible elements on a measure space. The reason for this is that in Isabelle, a function cannot be partial and it is necessary to define the measure function on every subset of $\Omega_\calM$; the choice of setting these measures to $0$ is arbitrary but does not entail any contradiction.


\newcommand{\calC}{\mathcal{C}}
\newcommand{\sigmeas}[1]{\Upsilon(#1)}
\newcommand{\isasigma}{\texttt{sigma}}

We can associate to any $\sigma$-algebra $\calC \subseteq 2^{\Omega}$ a measure space with a uniformly null measure: $\sigmeas{\calC} \isdef (\Omega, \calC, (\lambda\, x. 0))$. In Isabelle, this measure space is denoted by $\isasigma\ \Omega\ \calC$.

A function between two measurable spaces is \emph{measurable} if the preimage of every measurable set is measurable. 
In Isabelle, sets of measurable functions are defined as follow:
\[\begin{array}{lcl}
\measurable & :: & \alpha\,\measure \rightarrow \beta\,\measure\rightarrow \left(\alpha \rightarrow \beta\right) \isaset\\
\measurable\ \calM\ \calN\ \mu & =&
\setof{f: \Omega_\calM \rightarrow \Omega_\calN}{\forall A\in \calA_\calN.\, f^{-1}(A) \cap \Omega_\calM \in \calA_M}
\end{array}\]

\newcommand{\probspace}{\texttt{prob-space}}
\newcommand{\finitemeasure}{\texttt{finite-measure}}
\newcommand{\AEv}{\textsc{AE}}

Probability measures are measure spaces on which the measure of $\Omega$ is finite and equal to 
$1$. 
In Isabelle, they are defined by a \emph{locale}; this allows one to delimit a range in which the existence of a measure satisfying the desired assumptions is assumed, instead of having to explicitly add the corresponding hypotheses in every theorem, which would be tedious.

\[\begin{array}{l}
\keyw{locale}\ \probspace =  \finitemeasure\ +\keyw{assumes}\ \mu_\calM(\Omega_\calM)\ =\ 1 
\end{array}\]
A \emph{random variable} on a probability space $\calM$ is a measurable function with domain $\Omega_\calM$.
Collections of random variables are called \emph{stochastic processes}. In most cases, stochastic processes are indexed by a totally ordered set, representing time, such as $\dN$ or $\dR^+$.
In what follows, we will consider properties that hold \emph{almost surely} (or \emph{almost everywhere}), {i.e.,} are such that the elements for which they do not hold reside within a set of measure $0$:
\begin{align*}
\keyw{lemma}\ \textsc{AE-iff}& :\\\
(\AEv_\calM\,x.\ P\ x)& \Leftrightarrow (\exists N\in \calA_\calM.\, \mu_\calM(N) = 0 \wedge \setof{x}{\neg P\ x}\subseteq N)
\end{align*}

Given measure spaces $\calM$ and $\calN$, we say that $\calN$ is a \emph{subalgebra} of $\calM$ if $\Omega_\calM = \Omega_\calN$ and $\calA_\calN\subseteq \calA_\calM$.

\section{Modeling equity markets in discrete time}

\label{sect:equity_market}

\subsection{General definitions}

\label{sect:general_def}

\newcommand{\univ}{\texttt{UNIV}}

An equity market is characterized by the set of assets that can be traded and the price at which they are traded\footnote{This is a simplification as in practice, two prices are associated with each asset: a \emph{bid price}, which represents the price traders are willing to pay to buy the asset, and an \emph{ask} price, which represents the price traders are willing to sell the asset for. Bid prices are always lower than ask prices, but on markets on which high volumes of assets are traded, both prices are very close.}. A subset of these assets represents the basic securities that can be traded, these are the \emph{stocks}. Examples of stocks are shares on companies like Google, Apple, Facebook or Amazon, which can be traded on the stock market. The remaining assets are viewed as \emph{derivative products}, the value of which typically depends on that of some stocks. Examples of derivative products are \emph{futures} on Facebook, or \emph{basket options} on Apple and Google. Their precise definition is not important at this point, these are assets with a value depending on that of one or several stocks. The price at which an asset can be traded at each time is a random variable, this price is thus represented by a stochastic process; and in this case for which time is discrete, these stochastic processes are indexed by $\dN$. At time $n$, the random variable associated with an asset thus represents the value of this asset on time interval $[n, n+1[$. Note that in this general setting, there is no relationship between the price processes of assets and that of stocks. 
As we are concerned with computing fair prices for derivative products, equity markets are defined in such a way that there always exists at least one derivative product. 
\newcommand{\discrmkt}{\texttt{discr-mkt}}
\newcommand{\discmarket}{\texttt{discrete-market}}
\[\begin{array}{lcl}
	\discrmkt & :: & \beta\ \isaset \rightarrow (\beta \rightarrow(\dN \rightarrow \alpha \rightarrow\dR)) \rightarrow\dB\\
	\discrmkt\ S\ P & \Leftrightarrow & S \neq \univ
\end{array}\]
Equity markets are defined as a type, from which the stocks and prices can be obtained: 
\[\begin{array}{l}
\keyw{typedef} (\alpha, \beta)\, \discmarket = \setof{(S,P)}{\discrmkt\ S\ P} 
\end{array}\]

\newcommand{\stocks}{\texttt{stocks}}
\newcommand{\mkt}{\texttt{Mkt}}
\newcommand{\fst}{\texttt{fst}}
\newcommand{\snd}{\texttt{snd}}
\newcommand{\rep}{\texttt{Rep}}
\newcommand{\prices}{\texttt{prices}}

\[\begin{array}{lcl}
	\stocks & :: & (\alpha,\beta)\, \discmarket \rightarrow \beta\ \isaset\\
	\stocks\ \mkt &= & \fst\ (\rep\text{-}\discmarket\ \mkt)
\end{array}\]

\[\begin{array}{lcl}
\prices & :: & (\alpha,\beta)\, \discmarket \rightarrow \beta\rightarrow \dN \rightarrow \alpha\rightarrow \dR\\
\prices\ \mkt &= & \snd\ (\rep\text{-}\discmarket\ \mkt)
\end{array}\]

We next consider \emph{quantity processes}. These are used to represent the fact that assets can be bought and sold; in particular, it is possible on financial markets to sell an asset that is not withheld: when this occurs, we say the seller is \emph{short on the asset} and owns a negative amount of the asset. When {the holder owns a positive amount of the asset, we say the {holder} is \emph{long on the asset}. 
 We assume that any portion of the asset may be traded, thus the quantity withheld is a real number.
Quantity processes are  formalized as functions that associate a stochastic process to each asset. By convention, for $n>0$, if $q$ is a quantity process and $a$ is an asset, then $q\ a\ n\ w$ represents the quantity (positive if we are long the asset and negative if we are short the asset) of asset $a$ withheld on the time interval $]n-1,n]$ for scenario  $w$. The value of a quantity process at time $0$ is thus unimportant. Intuitively, the reason for such a convention is that, at time $n$, a quantity process is meant to only depend on the information available up to time $n-1$. More formally, in both discrete and continuous-time models, quantity processes of interest will be required to be \emph{predictable processes}, and the convention on quantity processes allows for a uniform presentation.
We define operators that permit to construct and combine quantity processes.

\newcommand{\qtyempty}{\texttt{qty-empty}}
\newcommand{\qtysingle}{\texttt{qty-single}}
\newcommand{\qtysum}{\texttt{qty-sum}} 
\newcommand{\qtymult}{\texttt{qty-mult-comp}} 
\newcommand{\qtyrem}{\texttt{qty-rem-comp}} 
\newcommand{\qtyrepl}{\texttt{qty-repl-comp}} 
\newcommand{\asset}{\texttt{asset}}
\newcommand{\qty}{\texttt{prc}}
\newcommand{\prd}{\texttt{prd}}

\[\begin{array}{lcl}
\qtyempty & :: & \beta\rightarrow \dN \rightarrow \alpha\rightarrow \dR\\
\qtyempty &= & (\lambda x\ n\ w.\ 0)
\end{array}\]

\[\begin{array}{lcl}
\qtysingle & :: & \beta \rightarrow (\dN \rightarrow \alpha\rightarrow \dR) \rightarrow \beta\rightarrow \dN \rightarrow \alpha\rightarrow \dR\\
\qtysingle\ \asset\ \qty &= & \qtyempty(\asset := \qty)
\end{array}\]

\[\begin{array}{lcl}
\qtysum & :: & (\beta\rightarrow \dN \rightarrow \alpha\rightarrow \dR) \rightarrow (\beta\rightarrow \dN \rightarrow \alpha\rightarrow \dR) \rightarrow\\
&&\quad \beta\rightarrow \dN \rightarrow \alpha\rightarrow \dR\\
\qtysum\ q_1\ q_2 &= & (\lambda x\ n\ w.\ (q_1\ x\ n\ w) + (q_2\ x\ n\ w))
\end{array}\]

\[\begin{array}{lcl}
\qtymult & :: & (\beta\rightarrow \dN \rightarrow \alpha\rightarrow \dR) \rightarrow (\dN \rightarrow \alpha\rightarrow \dR) \rightarrow\\
&&\quad \beta\rightarrow \dN \rightarrow \alpha\rightarrow \dR\\
\qtymult\ q\ \prd &= & (\lambda x\ n\ w.\ (q\ x\ n\ w).(\prd\ n\ w))
\end{array}\]

\[\begin{array}{lcl}
\qtyrem & :: & (\beta\rightarrow \dN \rightarrow \alpha\rightarrow \dR) \rightarrow (\dN \rightarrow \alpha\rightarrow \dR) \rightarrow\\
&&\quad \beta\rightarrow \dN \rightarrow \alpha\rightarrow \dR\\
\qtyrem\ q\ \asset &= & q(\asset := (\lambda n\ w. 0))
\end{array}\]

Intuitively, {\qtyempty} represents the quantity process in which no asset is bought or sold, and {\qtysingle} is the process for which a single asset is potentially bought or sold\footnote{Recall that in Isabelle, the notation $f(a:=b)$ represents an update of function $f$ so that the image of $a$ becomes $b$}. The other operators permit to respectively add quantity processes, to multiply all of them by another process, and to nullify the quantity of an asset.

Related to the notion of a quantity process is that of its support set, which consists of all the assets that are potentially bought or sold at some point for some scenario. This leads to the definition of portfolios, which are quantity processes that admit a finite support set. Stock portfolios are portfolios for which the support set consists only of stocks.

\newcommand{\supportset}{\texttt{support-set}}
\newcommand{\portfolio}{\texttt{portfolio}}
\newcommand{\finite}{\texttt{finite}}
\newcommand{\stockpf}{\texttt{stock-portfolio}}

\[\begin{array}{lcl}
\supportset & :: & (\beta\rightarrow \dN \rightarrow \alpha\rightarrow \dR) \rightarrow \beta\ \isaset\\
\supportset\ q &= & \setof{a}{\exists n\,w.\ q\ a\ n\ w \neq 0}
\end{array}\]

\[\begin{array}{lcl}
\portfolio & :: & (\beta\rightarrow \dN \rightarrow \alpha\rightarrow \dR) \rightarrow \dB\\
\portfolio\ p &\Leftrightarrow & \finite\ (\supportset\ p)
\end{array}\]

\[\begin{array}{lcl}
\stockpf & :: & (\alpha,\beta)\, \discmarket\rightarrow (\beta\rightarrow \dN \rightarrow \alpha\rightarrow \dR) \rightarrow \dB\\
\stockpf\ p &\Leftrightarrow & \portfolio\ p \wedge \supportset\ p \subseteq \stocks\ \mkt
\end{array}\]

\newcommand{\apple}{\mathrm{Apl}}
\newcommand{\fbk}{\mathrm{Fbk}}
\newcommand{\goog}{\mathrm{Goog}}

\begin{example}\label{ex:stockpf}
	Consider a market $\mkt$ with stocks including shares on Apple, Facebook and Google: $\set{\apple,\, \fbk,\, \goog} \subseteq \stocks\ \mkt$. We can construct the following portfolio 
	\[p_1\isdef \qtysum\ (\qtysingle\ \apple\ (\lambda\ n\ w. n))\ (\qtysingle\ \goog\ (\lambda\ n\ w. -n)).\]
	This is portfolio in which we are long $n$ shares of Apple and short $n$ shares of Google until time $n$ for all scenarios; it has a support set consisting of Apple and Google and is thus a stock portfolio.
	\[\begin{array}{|c|c|c|c|c|c|}
	\hline		
	\textrm{Time}  & 1 & 2 & 3 & 4\\
	\hline
	\hline
	\apple\ \text{quantity}  & 1 & 2 & 3 & 4  \\	
	\goog\ \text{quantity}  & -1 & -2 & -3 & -4  \\
	\hline
	\end{array}\]
\end{example}

\newcommand{\updval}{closing value}
\newcommand{\updvalproc}{{\updval} process}

We now define \emph{value processes} and \emph{\updvalproc} for portfolios. Intuitively, the value process of a portfolio at time $n$ represents the total amount of cash that is necessary to invest in the assets of the portfolio until time $n+1$, and the {\updvalproc} of a portfolio at time $n$ represents the total amount of cash received/owed when closing out all positions at time $n$. 
The {\updvalproc} of a portfolio at time $0$ can be defined arbitrarily; a standard practice consists in setting its value to that of the value process of the portfolio at time $0$. 
	Note that if the composition of the portfolio does not change between 
times $]n-1,n]$ and $]n,n+1]$, then the value of the {\updvalproc} at time $n$ is the same as that of the value process.

\newcommand{\valproc}{\texttt{val-process}}
\newcommand{\tmpup}{\texttt{tmp-cl-val}}
\newcommand{\upvp}{\texttt{cls-val-process}}

\[\begin{array}{lcl}
\valproc & :: & (\alpha,\beta)\, \discmarket\rightarrow (\beta\rightarrow \dN \rightarrow \alpha\rightarrow \dR) \rightarrow\\
& & \quad \dN\rightarrow \alpha \rightarrow \dR\\
\valproc\ \mkt\ p &= & \keyw{if }\ \neg(\portfolio\ p)\ \keyw{ then }\ (\lambda n\ w.\ 0)\ \keyw{ else }\\
 \quad\quad\left(\lambda n\ w.\ \sum_{a \in \supportset\ p} ((\prices\ \mkt)\ a\ n\ w) * (p\ a\ (n+1)\ w)\right)\span\span
\end{array}\]

\[\begin{array}{lcl}
\tmpup & :: & (\alpha,\beta)\, \discmarket\rightarrow (\beta\rightarrow \dN \rightarrow \alpha\rightarrow \dR) \rightarrow\\
& & \quad \dN\rightarrow \alpha \rightarrow \dR\\
\tmpup\ \mkt\ p\ 0&= & \valproc\ \mkt\ p\ 0\\
\tmpup\ \mkt\ p\ (n+1) & = &\\
\quad\quad \left(\lambda w.\ \sum_{a \in \supportset\ p} ((\prices\ \mkt)\ a\ (n+1)\ w) * (p\ a\ (n+1)\ w)\right)\span\span
\end{array}\]

\[\begin{array}{lcl}
\upvp & :: & (\alpha,\beta)\, \discmarket\rightarrow (\beta\rightarrow \dN \rightarrow \alpha\rightarrow \dR) \rightarrow\\
& & \quad \dN\rightarrow \alpha \rightarrow \dR\\
\upvp\ \mkt\ p &= & \keyw{if }\ \neg(\portfolio\ p)\ \keyw{ then }\ (\lambda n\ w.\ 0)\ \keyw{ else }\\
\quad\quad\left(\lambda n\ w.\ \tmpup\ \mkt\ p\ n\ w\right)\span\span
\end{array}\]

\begin{example}
	Assume the Apple and Google shares have deterministic prices given by the table below. Then the value process and {\updvalproc} of portfolio $p_1$ defined in Example \ref{ex:stockpf} are given in the same table. 
	\begin{center}
		\[\begin{array}{|c|c|c|c|c|}
		\hline		
		\textrm{Time} & 0 & 1 & 2 & 3\\
		\hline
		\hline
		\apple\ \text{quantity}  & - & 1 & 2 & 3  \\	
		\goog\ \text{quantity}  & - & -1 & -2 & -3  \\
		\hline
		\apple \text{ value} & 100 & 98 & 96 & 98 \\	
		\goog \text{ value} & 90 & 92 & 98 & 95.5 \\
		\hline 
		\valproc\ \mkt\ p_1 & 10 & 12 & -6 & 10\\
		\upvp\ \mkt\ p_1 & 10 & 6 & -4 & 7.5\\
		\hline
		\end{array}\]
	\end{center}
\end{example}

\emph{Self-financing portfolios} are portfolios in which no cash is invested except possibly at inception. 
A portfolio is self-financing if its {\updval} and value at time $n+1$ are identical;
this means that the value of the portfolio may be affected by the evolution of the market but not by the changes in its composition.

\newcommand{\selffin}{\texttt{self-financing}}
\newcommand{\mkselffin}{\texttt{self-fin}}

\[\begin{array}{lcl}
\selffin & :: & (\alpha,\beta)\, \discmarket\rightarrow (\beta\rightarrow \dN \rightarrow \alpha\rightarrow \dR)\\
&& \quad \rightarrow \dB\\
\selffin\ \mkt\ p& \Leftrightarrow & \\
\quad \quad \forall n.\, \valproc\ \mkt\ p\ (n+1) = \upvp\ \mkt\ p\ (n+1)\span\span
\end{array}\]

A self-financing portfolio with initial value $v_0$ can be obtained starting from an arbitrary portfolio, provided the market contains an asset that never admits a price equal to 0, by buying (resp. selling) the required quantity of the asset with the extra (resp. missing) cash. 


\begin{example}
	Portfolio $p_1$ of Example \ref{ex:stockpf} is not self-financing. Assume the stock price of Facebook is deterministic and given in the table below.
	\[\begin{array}{|c|c|c|c|c|}
	\hline		
	\textrm{Time} & 0 & 1 & 2 & 3\\
	\hline
	\hline
	\fbk & 5 & 4 & 4 & 5 \\	
	\hline
	\end{array}\]	
	Then we can construct a self-financing portfolio $p_2$ with initial value $0$, that has the same quantity processes as $p_1$ for Apple and Google. The quantities of stocks and the value and {\updvalproc}es of $p_2$ are given in the following table.
	\[\begin{array}{|c|c|c|c|c|c|}
	\hline		
	\textrm{Time} & 0 & 1 & 2 & 3 & 4\\
	\hline
	\hline
	\apple\ \text{quantity} & - & 1 & 2 & 3 & 4  \\	
	\goog\ \text{quantity} & - & -1 & -2 & -3 & -4 \\
	\fbk\ \text{quantity} & - & -2 & -3.5 & -3 & -3.5 \\	
	\hline 
	\apple \text{ value} & 100 & 98 & 96 & 98 & - \\	
	\goog \text{ value} & 90 & 92 & 98 & 95.5 & - \\
	\fbk \text{ value} & 5 & 4 & 4 & 5 & - \\	
	\hline 
	\valproc\ \mkt\ p_2 & 0 & -2 & -18 & -7.5 & -\\
	\upvp\ \mkt\ p_2 & 0 & -2 & -18 & -7.5 & -\\
	\hline
	\end{array}\]
	The value process of $p_2$ at time 3 is computed by having a quantity $-3.5$ of Facebook stock in the portfolio until time 4.
	{For instance, at time $0$, the holder buys a share of Apple for 100\euro{} and sells a (borrowed) share of Google for 90\euro{}, creating a portfolio for a total cost of $10$\euro{}. To make the portfolio self-financed with initial value 0, this cost is compensated by 
	selling $2$ (borrowed) shares of Facebook at 5\euro{} each, the total cost of the created portfolio is then $0$\euro{}.}
\end{example}

\subsection{Modeling time-dependent information}

Filtrations are used to represent information accumulated over time. Formally, they are defined as a collection of increasing subalgebras over a totally ordered set with a minimal element $\bot$ --typically $\dN$ or $\dR^+$. 
\newcommand{\filtration}{\texttt{filtration}}
\newcommand{\calF}{\mathcal{F}}
\newcommand{\linorder}{\texttt{linorder}}
\newcommand{\linorderbot}{\texttt{linorder-bot}}
\newcommand{\botcl}{\texttt{bot}}
\newcommand{\subalgebra}{\texttt{subalgebra}}

\[\keyw{class}\ \linorderbot = \linorder + \botcl \]

\[\begin{array}{lcl}
\filtration & :: & \alpha\,\measure \rightarrow ((\iota::\linorderbot) \rightarrow \alpha\,\measure) \rightarrow \dB\\
\filtration\ \calM\ \calF & \Leftrightarrow& (\forall t.\ \subalgebra\ \calM\ \calF_t)\ \wedge\\
&& (\forall s\ t.\ s\leq t\Rightarrow\  \subalgebra\ \calF_{t}\ \calF_s)
\end{array}\]

\newcommand{\inittrivfilt}{initially trivial filtration}
\newcommand{\inittriv}{\texttt{init-triv-filt}}

In general, when a filtration $\calF$ representing available information is provided, we will mainly be interested in stochastic processes that depend on this information. There are two categories of such stochastic processes of interest for our purpose: \emph{adapted stochastic process}, that at time $n$ are $\calF_n$-measurable; and \emph{predictable stochastic processes}, that at time $n >0$ are $\calF_{n-1}$-measurable. The definition of adapted stochastic processes in the more general case is a straightforward generalization of that in the discrete case, which is the one that is formalized below. We also introduce abbreviations for stochastic processes with a range in a borel measure space.

\newcommand{\adaptdsp}{\texttt{adapt-sp}}
\newcommand{\boreladapt}{\texttt{borel-adapt-sp}}
\newcommand{\borel}{\texttt{borel}}

\[\begin{array}{lcl}
\adaptdsp & :: & (\iota \rightarrow \alpha\,\measure) \rightarrow (\iota \rightarrow \alpha \rightarrow \beta)\rightarrow\\ & & \quad \beta\,\measure \rightarrow \dB\\
\adaptdsp\ \calF\ X\ \calN\ & \Leftrightarrow & \forall t.\ X_t \in \measurable\ \calF_t\ \calN
\end{array}\]

\[\begin{array}{rcl}
\keyw{abbreviation}\ \boreladapt\ \calF\ X\ \equiv\ \adaptdsp\ \calF\ X\ \borel
\end{array}\]


\newcommand{\predictsp}{\texttt{predict-sp}}
\newcommand{\borelpredict}{\texttt{borel-predict-sp}}

\[\begin{array}{lcl}
\predictsp & :: & (\dN \rightarrow \alpha\,\measure) \rightarrow (\dN \rightarrow \alpha \rightarrow \beta)\rightarrow\\ & & \quad \beta\,\measure \rightarrow \dB\\
\predictsp\ \calF\ X\ \calN\ & \Leftrightarrow & X_0 \in \measurable\ \calF_0\ \calN\wedge\\
&&\forall n.\ X_{n+1} \in \measurable\ \calF_n\ \calN
\end{array}\]

\[\begin{array}{rcl}
\keyw{abbreviation}\ \borelpredict\ \calF\ X\ \equiv\ \predictsp\ \calF\ X\ \borel
\end{array}\]

In our context, filtrations are meant to represent the currently available information. A standard filtration used in financial mathematics is the one defined as follows: for all $n\geq 0$, $\calF_n$ is the smallest subalgebra of $\calM$ in which for any stock $s$ and time $k\leq n$, the price process $(\prices\ \mkt)\ s\ k$ is borel-measurable. It is straightforward to verify that $\calF$ is indeed a filtration.
In particular, at time $0$, there is no information available, thus the measure space $\calF_0 = \calF_\bot$ is trivial. Filtrations satisfying such a requirement are called \emph{{\inittrivfilt}s}. 

\[\begin{array}{lcl}
\inittriv & :: & \alpha\,\measure \rightarrow (\iota \rightarrow \alpha\,\measure) \rightarrow \dB\\
\inittriv\ \calM\ \calF & \Leftrightarrow& \filtration\ \calM\ \calF \wedge \sets\ \calF_\bot = \set{\emptyset, \Omega_\calM}
\end{array}\]

We define a locale for discrete equity markets by fixing a market and considering a probability space equipped with an arbitrary filtration that is initially trivial. 

\newcommand{\initrivpb}{\texttt{init-triv-prob-space}}
\newcommand{\deqm}{\texttt{disc-equity-market}}

\[\begin{array}{l}
	\keyw{locale}\ \initrivpb\ =\ \probspace\ +\\
	\quad \keyw{fixes}\ \calF::\dN\rightarrow (\alpha\, \measure)\\
	\quad \keyw{assumes}\ \inittriv\ \calF\\
	\keyw{locale}\ \deqm\ =\ \initrivpb\ +\\
	\quad \keyw{fixes}\ \mkt::(\alpha,\beta)\, \discmarket
\end{array}\]

Most of the assets that we will be considering in this locale are those that have an adapted price process w.r.t. the given filtration. Quantity processes in which only assets with an adapted price process are bought or sold are called \emph{support-adapted} quantity processes.

\newcommand{\supportadapt}{\texttt{support-adapt}}

\[\begin{array}{lcl}
\supportadapt & :: & (\alpha,\beta)\,\discmarket \rightarrow (\beta\rightarrow\dN \rightarrow\alpha\rightarrow\dR)\\ &&\quad \rightarrow \dB\\
\supportadapt\ \mkt\ \qty & \Leftrightarrow& \forall a\in \supportset\ p.\\
&&\quad \boreladapt\ \calF\ (\prices\ \mkt\ a)
\end{array}\]

The portfolios that can reasonably be constructed are those for which the amounts that are bought or sold of each asset on the time interval $]n, n+1]$ is known at time $n$. In other words, these are portfolios for which the quantity of each asset is a predictable process; such portfolios are called \emph{trading strategies}.

\newcommand{\tradingstrat}{\texttt{trading-strat}}

\[\begin{array}{lcl}
\tradingstrat & :: & (\beta \rightarrow \dN\rightarrow \alpha\rightarrow\dR) \rightarrow \dB\\
\tradingstrat\ p & \Leftrightarrow& \portfolio\ p\ \wedge\\
& & \quad  (\forall a\in \supportset\ p.\ \borelpredict\ \calF\ (p\ a))
\end{array}\]

In particular, the value process of a support-adapted trading strategy is itself an adapted process:

\[\begin{array}{l}
\keyw{lemma}\  \textsc{trading-strategy-adapted}\\
\quad \keyw{assumes}\ \tradingstrat\ p\\
\quad \keyw{and}\ \supportadapt\ \mkt\ p\\
\quad \keyw{shows}\ \boreladapt\ \calF\ (\valproc\ \mkt\ p)
\end{array}\]
Since the filtration $\calF$ is assumed to be initially trivial, such a strategy necessarily admits a constant value at inception:
\[\begin{array}{l}
\keyw{lemma}\  \textsc{trading-strategy-init}\\
\quad \keyw{assumes}\ \tradingstrat\ p\\
\quad \keyw{and}\ \supportadapt\ \mkt\ p\\
\quad \keyw{shows}\ \exists c.\ \forall \omega\in \Omega_\calM.\ \valproc\ \mkt\ p\ 0\ w = c
\end{array}\]
\newcommand{\initval}{\texttt{init-value}}
We denote by $\initval\ p$ the constant value equal to the value process of a trading strategy at time 0.

\section{The notion of a fair price}\label{sect:fair}

\subsection{Definitions}

We define the notion of a fair price, which is meant to represent the price at which a derivative should be bought or sold. Intuitively, a fair price for an asset is one that does not allow a buyer or seller of the asset of making a risk-free profit thanks to this transaction. Making a risk-free profit is called an \emph{arbitrage}. We begin by formally defining the notion of an arbitrage. This is a self-financing trading strategy with a zero initial value, that at some point in time is almost surely positive and with a strictly positive probability of making a gain.
Although such arbitrage opportunities do exist in real financial markets, they are generally quickly exploited and disappear: in fact, there is an entire category of traders on markets with the goal of detecting and exploiting arbitrages as quickly as possible. Pricing results in financial mathematics are based on a no-arbitrage assumption. 
\hide{\Mnacho{todo: develop}
 
 \Niko{Cela n'a pas d'impact sur le papier, mais je me demandais si on pouvait dans la définition faire dépendre $n$ de $\omega$ (ie garantir qu'on fait un profit à un moment dans le futur, le moment dépendant de l'alea).}}
 
 \newcommand{\arbitrage}{\texttt{arbitrage-process}}
\newcommand{\calP}{\mathcal{P}}
\newcommand{\viable}{\texttt{viable-market}}
\[\begin{array}{lcl}
\arbitrage  & :: & (\alpha,\beta)\,\discmarket \rightarrow (\beta\rightarrow \dN\rightarrow\alpha\rightarrow \dR) \rightarrow \dB\\
\arbitrage\ \mkt\ p & \Leftrightarrow &\\
\quad (\exists m\in \dN.\span\span\\
\quad\quad (\tradingstrat\ p)\ \wedge\
(\selffin\ p)\ \wedge\span\span\\
\quad\quad (\forall \omega \in \Omega_\calM.\ (\valproc\ p)\ 0\ \omega = 0)\ \wedge\span\span\\
\quad\quad (\AEv_\calM\ \omega.\ (\upvp\ p)\ m\ \omega \geq 0)\ \wedge\span\span\\
\quad\quad (\calP(\setof{\omega\in \Omega_\calM}{(\upvp\ p)\ m\ \omega > 0}) > 0))\span\span\\
\end{array}\]

Next we define the notion of a \emph{price structure} for a derivative. Derivatives are characterized by their maturity and the payoff they deliver at maturity; a price structure is a stochastic process with a constant initial value that coincides with the payoff of the derivative almost everywhere at maturity. The initial value of a price structure will represent the price of the derivative under consideration.

\newcommand{\prstruct}{\texttt{price-struct}}
\newcommand{\pyf}{\kappa}
\newcommand{\pr}{\texttt{pr}}
\[\begin{array}{lcl}
\prstruct & :: & (\alpha\rightarrow\dR)\rightarrow\dN \rightarrow \dR \rightarrow (\dN\rightarrow\alpha \rightarrow \dR) \rightarrow \dB\\
\prstruct\ \pyf\ T\ \pi\ \pr & \Leftrightarrow& (\forall \omega \in \Omega_\calM.\ \pr\ 0\ \omega = \pi)\ \wedge\\
& & (\AEv_\calM\ \omega.\ \pr\ T\ \omega = \pyf\ \omega)\ \wedge\\
& & (\pr\ T \in \borelmeasurable\ \calF_T)
\end{array}\]

In order to formalize the notion of a fair price for a derivative, we need to formalize the fact that buying or selling the derivative at that price $\pi$ does not lead to any arbitrage opportunity. More precisely, it should not be possible to obtain an arbitrage process using only stocks from the market and an asset with a price process identical to a price structure of the derivative, with an initial value $\pi$. In order to guarantee the existence of such an asset, we define the notion of coincidence between two markets.

\newcommand{\coincideson}{\texttt{coincides}}
\newcommand{\fairprice}{\texttt{fair-price}}

\[\begin{array}{lcl}
\coincideson & :: & (\alpha,\beta)\,\discmarket \rightarrow (\alpha,\beta)\,\discmarket \rightarrow\\
& & \quad \beta\ \isaset \rightarrow\dB\\
\coincideson\ \mkt\ \mkt'\ A & \Leftrightarrow& \stocks\ \mkt = \stocks\ \mkt'\ \wedge\\
& & \forall x.\ x\in A\Rightarrow \prices\ \mkt\ a = \prices\ \mkt'\ a
\end{array}\]

\[\begin{array}{lcl}
\fairprice & :: & (\alpha,\beta)\,\discmarket \rightarrow \dR\rightarrow (\alpha\rightarrow \dR) \rightarrow\\
& & \quad \dN \rightarrow\dB\\
\fairprice\ \mkt\ \pi\ \pyf\ T & \Leftrightarrow& (\exists\pr.\ \prstruct\ \pyf\ T\ \pi\ \pr \wedge\\
& & \quad (\forall a\ \mkt'\ p.\ a\notin \stocks\ \mkt \Rightarrow\\
& & \quad\quad \quad(\coincideson\ \mkt\ \mkt')\ \wedge\\
& & \quad\quad \quad (\prices\ \mkt'\ a = \pr)\ \wedge\\
& & \quad\quad\quad (\portfolio\ p)\ \wedge\\
& & \quad\quad\quad (\supportset\ p\subseteq \stocks\ \mkt \cup \set{a}) \Rightarrow\\
& & \quad\neg\arbitrage\ \mkt'\ p))
\end{array}\]

\subsection{Replicating portfolios}

\label{sect:fair_price_unique}

We prove the central result that, under the hypothesis that a \emph{replicating portfolio} exists for a given derivative, the latter admits a fair price that is unique. A replicating portfolio for a given derivative is a self-financing trading strategy that consists of stocks only, and that at maturity, has a value identical to the payoff of the derivative almost everywhere. If such a portfolio exists, then the derivative is \emph{attainable}, and if every derivative available on a market is attainable, then the market is \emph{complete}:
\newcommand{\attainable}{\texttt{attainable}}
\newcommand{\repl}{\texttt{replic-pf}}
\newcommand{\complete}{\texttt{complete-market}}


\[\begin{array}{lcl}
\repl  & :: & (\beta\rightarrow \dN\rightarrow \alpha\rightarrow \dR) \rightarrow (\alpha\rightarrow \dR) \rightarrow\ \dN \rightarrow\dB\\
\repl\ p\ \pyf\ T & \Leftrightarrow & (\stockpf\ \mkt\ \ p)\ \wedge\\
& & (\selffin\ p)\,\wedge  (\tradingstrat\ p)\, \wedge\\
& &  (\AEv_\calM\ \omega.\ \upvp\ \mkt\  p\ T\ \omega = \pyf\ w)
\end{array}\]
\[\begin{array}{lcl}
\attainable  & :: & (\alpha \rightarrow \dR) \rightarrow \dN \rightarrow\dB\\
\attainable\ \pyf\ T & \Leftrightarrow & (\exists p.\ \repl\ p\ \pyf\ T)
\end{array}\]
\[\begin{array}{lcl}
\complete  & :: & \dB\\
\complete & \Leftrightarrow & \forall T.\ \forall \pyf\in (\borelmeasurable\ \calF_T).\ \attainable\ \pyf\ T
\end{array}\]

The existence of a replicating portfolio by itself is not sufficient to guarantee the existence of a fair price: indeed, if for example it is already possible to construct an arbitrage process on the market using only stocks, then there clearly cannot be any fair price for any derivative product. It is thus necessary to forbid arbitrage opportunities using only stocks from the market. This is captured by the notion of a \emph{viable market}.

\[\begin{array}{lcl}
\viable & :: & (\alpha,\beta)\, \discmarket \rightarrow \dB\\
\viable\ \mkt & \Leftrightarrow& \forall p.\ \stockpf\ p\ \Rightarrow\\
& &\quad \neg \arbitrage\ \mkt\ p
\end{array}\]

We obtain the following results.
{We first show that, if the market is viable, every derivative admitting a replicating portfolio has a fair price that is the initial
value of the replicating portfolio.}

\[\begin{array}{l}
\keyw{lemma}\  \textsc{replicating-fair-price}\\
\quad \keyw{assumes}\ \viable\ \mkt\\
\quad \keyw{and}\ \repl\ p\ \pyf\ T\\
\quad \keyw{and}\ \supportadapt\ \mkt\ p\\
\quad \keyw{shows}\ \fairprice\ \mkt\ (\initval\ p)\ \pyf\ T
\end{array}\]

We {then} provide a proof of the uniqueness of a fair price {for attainable derivatives} based on the existence of a stock on the market with a strictly positive price process. The proof could also be carried out assuming the existence of a stock on the market with a strictly negative price process, but that does not really make sense from a financial point of view. We also assume that the price processes of all stocks in the market are adapted to the filtration under consideration. 
\newcommand{\dmps}{\texttt{disc-mkt-pos-stock}}
\newcommand{\posstock}{\texttt{pos-stock}}
\[\begin{array}{l}
\keyw{locale}\ \dmps\ =\ \deqm\ +\\
\quad \keyw{fixes}\ \posstock::\beta\\
\quad \keyw{assumes}\ \posstock\in  \stocks\ \mkt\\
\quad \keyw{and}\ \forall\, n\ \omega.\ \prices\ \mkt\ \posstock\ n\ \omega > 0\\
\quad \keyw{and}\ \forall\, a\in \stocks\ \mkt.\ \boreladapt\ \calF\ (\prices\ \mkt\ a)
\end{array}\]

\[\begin{array}{l}
\keyw{lemma}\  \textsc{replicating-fair-price-unique}\\
\quad \keyw{assumes}\ \repl\ p\ \pyf\ T\\
\quad \keyw{and}\ \fairprice\ \mkt\ \pi\ \pyf\ T\\
\quad \keyw{shows}\ \pi = \ (\initval\ p)
\end{array}\]

\section{Risk-neutral probability spaces}\label{sect:riskneutr}

\subsection{Interest rates and discounted values}

We begin by defining the notion of interest rates. 
The existence of an interest rate is modeled by assuming that the market contains a stock with a deterministic return. The price process of this stock is parameterized by an interest rate $r$. In this setting, the interest rate is constant, although there exist 
more general models in which the interest rate can be time-dependent, and even stochastic. 
\hide{\Niko{c'est aussi en dehors du papier, mais je me demandais si le résultat était conservé dans ces deux cas}}
\newcommand{\discrfr}{\texttt{disc-rfr-proc}}
\[\begin{array}{lcl}
\discrfr & :: & \dR\rightarrow \dN\rightarrow \alpha\rightarrow \dR\\
\discrfr\ r\ 0\ \omega & =& 1\\
\discrfr\ r\ (n+1)\ \omega& = &(1+r).(\discrfr\ n\ \omega)
\end{array}\]
We call risk-free asset any asset $a$ such that $\prices\ \mkt\ a = \discrfr\ r$ for some rate $r$, and define a locale for a market containing a risk-free asset. 
\newcommand{\rfreeasset}{\texttt{risk-free-stock-market}}
\newcommand{\riskfreeasset}{\texttt{rf-asset}}

\[\begin{array}{l}
\keyw{locale}\ \rfreeasset\ =\ \deqm\ +\\
\quad \keyw{fixes}\ \riskfreeasset::\beta\\
\quad \keyw{and}\ r::\dR\\
\quad \keyw{assumes}\ -1 < r\\
\quad \keyw{and}\ \riskfreeasset \in \stocks\ \mkt\\
\quad \keyw{and}\ \prices\ \mkt\ \riskfreeasset = \discrfr\ r
\end{array}\]

 Having a risk-free asset as a stock in a market makes it possible to deposit (by buying the asset) or borrow (by shorting the asset) cash on this market.
\begin{example}
	Assume there is a risk-free asset with an annual rate of $2\%$ on the market. This means that buying $100${\euro} worth of the asset today will permit to obtain $102${\euro} by selling the asset in one year. Assuming the time lapse between times $n$ and $n+1$ is a day and there are 252 business days in one year, the daily rate $r$ in the definition of {\discrfr} then satisfies the equation $(1+r)^{252} = 1.02$, so we have $r\approx 7.85.10^{-5}$.
\end{example}

\begin{remark}
Observe that if the market is viable, then all risk-free assets must have the same rate.
Indeed, if there exist two risk-free assets with interest rates $r_1 < r_2$
then 
an arbitrage can be constructed: it suffices to buy 1 share of the second asset and 
sell  1 share of the first one. 
The initial investment is $1-1 = 0$, and at time $n$ the {\updval} of the portfolio 
is $(1+r_2)^n - (1+r_1)^n > 0$.
\end{remark}
We also define the \emph{discounted value} of a stochastic process. This notion is related to that of the present value of a future cash-flow, given an interest rate. 

\newcommand{\discountfactor}{\texttt{discount-factor}}
\newcommand{\discval}{\texttt{discounted-value}}
\newcommand{\inverse}{\texttt{inverse}}

\[\begin{array}{lcl}
\discountfactor & :: & \dR\rightarrow \dN\rightarrow \alpha\rightarrow \dR\\
\discountfactor\ r\ n\ \omega & =& \inverse\ (\discrfr\ r\ n\ \omega)
\end{array}\]

\[\begin{array}{lcl}
\discval & :: & \dR\rightarrow (\dN\rightarrow \alpha\rightarrow \dR) \rightarrow \dN \rightarrow \alpha \rightarrow\dR\\
\discval\ r\ X & =& \lambda\, n\ \omega.\ (\discountfactor\ r\ n\ \omega).(X_n\ \omega)
\end{array}\]

\begin{example}
	Assume we have a viable market that contains a risk-free asset with a rate of $2\%$ per year, and that the price of a share of Apple today is 95\euro.
	Consider a forward contract for buying a share of Apple stock at a strike price of $98$\euro{} in two years. The fair price for this contract is obtained by computing the discounted value of the strike and subtracting it from the current price of a share today. Here the discounted value of the strike is $98.(1+0.02)^{-2} \approx 94.19$, hence the fair price of this contract is $0.81$\euro. Indeed, this amount of money can be used to construct a replicating portfolio as follows.
	\begin{enumerate}
		\item Borrow $94.19$\euro{} today. 
		\item Use the cash, along with the $0.81$\euro{} received at the sale of the contract to buy a share of Apple stock today.
		\item Wait for two years. 
		\item Sell the share of Apple stock to the buyer of the forward contract for 98\euro.
		\item Use this to reimburse the cash that was borrowed at the start and is now worth $94.19.(1+0.02)^2 \approx98$\euro.
	\end{enumerate}
\end{example}


\subsection{Conditional expectations and martingales}

The results of {Section \ref{sect:fair_price_unique} show that when a replicating portfolio exists for a given derivative, the fair price for this derivative is unique and equal to the initial value of the portfolio. In this section we prove that this initial value can be computed without explicitly constructing any replicating portfolio under the hypothesis of the existence of a \emph{risk-neutral probability space}. From a financial point of view, assets carry different levels of risk, and the more risky an asset, the higher the return buyers will be expecting when investing in the asset; this additional return is called the \emph{market price of risk}. 
A risk-neutral probability space is meant to represent a world in which investors do not expect an increased return for a more risky asset: they are neutral to risk and expect the returns of all assets to be identical.

\newcommand{\condexp}[2]{\mathbb{E}\left[#1\mid#2\right]}
The expected returns of assets are modeled using the notion of \emph{conditional expectations}. A conditional expectation is meant to represent the best approximation of a random variable given the currently available information. Formally, a conditional expectation of a random variable $X$ given a measure space $\calN$ that is a subalgebra of $\calM$ is any random variable $X_\calN$ that is $\calN$-measurable, and such that for any set $N\in \calN$, 
\[\int_{N}X_\calN\mathrm{d}\mu_\calM\ =\ \int_{N}X\mathrm{d}\mu_\calM.\]
A conditional expectation of $X$ given $\calN$ always exists as long as $X$ is integrable, and is almost surely unique, meaning that two conditional expectations of $X$ given $\calN$ are identical almost everywhere. In what follows, we will therefore refer to \emph{the} conditional expectation of $X$ given $\calN$, and denote it by $\condexp{X}{\calN}$. Conditional expectations are already formalized in Isabelle. 

\newcommand{\martingale}{\texttt{martingale}}
\newcommand{\integrable}{\texttt{integrable}}
\newcommand{\realcond}{\texttt{real-cond-exp}}

Conditional expectations are used to define \emph{martingales}. Given a filtration $\calF$, these are stochastic processes $(X_t)_t$ such that for all $t\leq s$, $X_t$ is the best estimation of $X_s$ given the information $\calF_t$. In other words, $X_t$ and $\condexp{X_s}{\calF_t}$ are equal almost everywhere.

\[\begin{array}{lcl}
\martingale & :: & \alpha\,\measure \rightarrow (\iota \rightarrow \alpha\, \measure) \rightarrow (\iota \rightarrow \alpha \rightarrow \dR) \rightarrow \dB\\
\martingale\ \calM\ \calF\ X & \Leftrightarrow&\\ 
\quad(\filtration\ \calM\ \calF)\, \wedge\, (\boreladapt\ \calF\ X)\,\wedge\,
 (\forall t.\, \integrable\ \calM\ X_t)\,\wedge \span\span\\
\quad (\forall t\, s.\, t\leq s \Rightarrow (\AEv_\calM\ \omega.\, X_t\ w = \realcond\ \calM\ \calF_t\ X_s\ w))\span\span
\end{array}\]

Because the risk-free asset we defined has a deterministic price process with a constant return rate, it is straightforward to verify that the discounted value of this price process is constant, and is trivially a martingale. In a risk-neutral probability space, the martingale property holds for \emph{all} the stocks of the market:

\newcommand{\riskneutralprob}{\texttt{risk-neutral-prob-space}}

\[\begin{array}{lcl}
\riskneutralprob & :: & \alpha\, \measure\rightarrow \dB\\
\riskneutralprob\ \calN & \Leftrightarrow& \probspace\ \calN\ \wedge\\
\quad \forall a\in (\stocks\ \mkt).\ \martingale\ \calN\ \calF\ (\discval\ r\ (\prices\ \mkt\ a))\span\span
\end{array}\]

\newcommand{\filteqce}{filtration-equivalence}
\newcommand{\filteqt}{filtration-equivalent}
\newcommand{\filtequiv}{\texttt{filt-equiv}}

\subsection{Filtration-equivalence}

If there were no relationship whatsoever between a risk-neutral probability space and the actual probability space, the former would not be of much use. In general, both spaces are assumed to be \emph{equivalent}, meaning that they agree on the events that have a zero probability. It turns out that when there is a filtration associated with a probability space, this notion can be relaxed into that of \emph{\filteqce}, which is sufficient for our purpose.

\[\begin{array}{lcl}
\filtequiv & :: & (\iota \rightarrow \alpha\, \measure)\rightarrow \alpha\,\measure \rightarrow \alpha\,\measure\rightarrow \dB\\
\filtequiv\ \calF\ \calM\ \calN & \Leftrightarrow& \filtration\ \calM\ \calF \wedge \calA_\calM = \calA_\calN\ \wedge\\
\quad \forall i\ A.\ A \in \calA_{\calF_i} \Rightarrow (\mu_\calM(A) = 0 \Leftrightarrow \mu_\calN(A) = 0)\span\span
\end{array}\]

When probability spaces are {\filteqt}, almost everywhere properties propagate from one space to the other. In particular, a replicating portfolio for a derivative in one given probability space will necessarily be a replicating portfolio for the derivative in a {\filteqt} probability space, even if the probabilities assigned to different events may defer.

\[\begin{array}{l}
\keyw{lemma}\ \textsc{filt-equiv-borel-AE-eq}\\
\quad \keyw{assumes}\ \filtequiv\ \calF\ \calM\ \calN\\
\quad \keyw{and}\ f\in \borelmeasurable\ \calF_i\\
\quad \keyw{and}\ g\in \borelmeasurable\ \calF_i\\
\quad \keyw{and}\ \AEv_\calM\ \omega.\ f\ w= g\ w\\
\quad \keyw{shows}\ \AEv_\calN\ \omega.\ f\ w = g\ w 
\end{array}\]

Provided integrability properties are guaranteed for the assets of a self-financing trading strategy, the latter is a martingale in a {\filteqt} risk-neutral probability space.

\[\begin{array}{l}
\keyw{lemma}\ \textsc{self-fin-trad-strat-mart}\\
\quad \keyw{assumes}\ \filtequiv\ \calF\ \calM\ \calN\\
\quad \keyw{and}\ \riskneutralprob\ \calN\\
\quad \keyw{and}\ \tradingstrat\ p\\
\quad \keyw{and}\ \selffin\ \mkt\ p\\
\quad \keyw{and}\ \stockpf\ \mkt\ p\\
\quad \keyw{and}\ \forall n.\ \forall a\in \supportset\ p.\ \integrable\ \calN\\ 
\quad\quad\quad(\lambda\, \omega.\ (\prices\ \mkt\ a\ n\ \omega)(p\ a\ (n+1)\ \omega))\\
\quad \keyw{and}\ \forall n.\ \forall a\in \supportset\ p.\ \integrable\ \calN\\ 
\quad\quad\quad(\lambda\, \omega.\ (\prices\ \mkt\ a\ (n+1)\ \omega)(p\ a\ (n+1)\ \omega))\\
\quad \keyw{shows}\ \martingale\ \calN\ \calF\ (\discval\ r\ (\upvp\ \mkt\ p))
\end{array}\]

We obtain the following result, which in a viable market, provides an effective way of computing the fair price of an attainable derivative when a {\filteqt} risk-neutral probability space exists:

\newcommand{\expect}[1]{\mathbb{E}\left[#1\right]}

\[\begin{array}{l}
\keyw{lemma}\ \textsc{replicating-expectation}\\
\quad \keyw{assumes}\ \filtequiv\ \calF\ \calM\ \calN\\
\quad \keyw{and}\ \riskneutralprob\ \calN\\
\quad \keyw{and}\ \kappa\in \borelmeasurable\ \calF_T\\
\quad \keyw{and}\ \repl\ p\ \kappa\ T\\
\quad \keyw{and}\ \forall n.\ \forall a\in \supportset\ p.\ \integrable\ \calN\\ 
\quad\quad\quad(\lambda\, \omega.\ (\prices\ \mkt\ a\ n\ \omega)(p\ a\ (n+1)\ \omega))\\
\quad \keyw{and}\ \forall n.\ \forall a\in \supportset\ p.\ \integrable\ \calN\\ 
\quad\quad\quad(\lambda\, \omega.\ (\prices\ \mkt\ a\ (n+1)\ \omega)(p\ a\ (n+1)\ \omega))\\
\quad \keyw{and}\ \viable\ \mkt\\
\quad \keyw{and}\ \calA_{\calF_0} = \set{\set{}, \Omega_\calM}\\
\quad \keyw{shows}\ \fairprice\ \mkt\ \expect{\discval\ r\ \kappa\ T}\ \kappa\ T
\end{array}\]

\section{Fair prices in the Cox-Ross-Rubinstein model}\label{sect:CRR}

\newcommand{\measurepmf}{\texttt{measure-pmf}}
\newcommand{\pmf}{\texttt{pmf}}
\newcommand{\streamspace}{\texttt{stream-space}}
\newcommand{\stream}{\texttt{stream}}
\newcommand{\snth}{\mathit{nth}}
\newcommand{\bernstream}{\texttt{bernoulli-stream}}
\newcommand{\bernpmf}{\texttt{bernoulli-pmf}}
\newcommand{\infct}{\texttt{infinite-coin-toss-space}}

The CRR model is a discrete-time model consisting of a market in which there are two stocks, a risk-free asset and a risky one. At every time $n$, the risky asset can only move upward or downward with respective probabilities $p$ and $1-p$. This means that the evolution of the risky asset price can be modeled by tossing at each time $n$ a coin that lands on its head with a probability $p$, and having the price move upward at time $n+1$ exactly when the coin lands on its head. The evolution of this price is thus controlled by sequences of coin tosses. In most introductory textbooks on the CRR model, these sequences are finite as the results are presented for a given derivative with a finite maturity. We choose to consider infinite sequences---or streams---of coin tosses for the sake of generality. Since at time $n$ no event other than the outcome of the coin toss is required, this outcome can be represented by a Bernoulli distribution of parameter $p$. In Isabelle, because discrete probability distributions and probability mass functions are isomorphic, the type of probability mass functions are defined as a subtype of measures \cite{HolzlLT15}, along with an injective representation function $\measurepmf:: \alpha\ \pmf \rightarrow \alpha\ \measure$. The Bernoulli distribution is thus defined as $\measurepmf\ (\bernpmf\ p)$. The measure space for infinite sequences of independent coin tosses is isomorphic to the infinite product of Bernoulli distributions with the same parameter. In Isabelle, this measure space is defined using the function $\streamspace:: \alpha\ \measure\rightarrow (\alpha\ \stream)\ \measure$. 
The measure space thus defined is the smallest one in which the function $\snth:: \alpha\ \stream \rightarrow \dN\rightarrow \alpha$ such that $(\snth\ s\ n)$ is the $n$th element of stream $s$ is measurable \cite{HolzlJAR}. The measure spaces we consider are defined as follows:
\[\begin{array}{lcl}
\bernstream & :: & \dR \rightarrow (\dB\ \stream)\ \measure\\
\bernstream\ p & = &\streamspace\ (\measurepmf\ (\texttt{bernoulli-pmf}\ p))
\end{array}\]
We define a locale in which we impose that $0\leq p\leq 1$, and thus obtain a probability space:
\[\begin{array}{l}
\keyw{locale}\ \infct =  \\
{\keyw{fixes}\ p\ \keyw{and}\ M}\\
\keyw{assumes}\ 0\leq p\leq 1\ \keyw{and}\ M = \bernstream\ p\\
\end{array}\]

\newcommand{\natfilt}{\calF^{\mathrm{nat}}}
\newcommand{\projt}{\pi^{\top}}
\newcommand{\infiltr}{\texttt{infinite-cts-filtration}}

The maximal amount of information that should be available at time $n$ is the outcome of the first $n$ coin tosses, and we define a filtration $\natfilt$ accordingly: intuitively, in this filtration, two streams of coin tosses with the same first $n$ outcomes cannot occur in distinct sets that are measurable in $\natfilt$. In our setting, each restricted measure space $\natfilt_n$ can be defined as generated by an arbitrary measurable function which maps all streams that agree on the first $n$ coin tosses to the same element. We thus considered the sequence of so-called \emph{pseudo-projection functions} $(\projt_n)_{n\in \dN}$, where:
\[\begin{array}{ccccl}
\projt_n & : & \Omega_\calM & \rightarrow & \Omega_\calM\\
& & (w_1,\cdots, w_n, w_{n+1},\cdots) & \mapsto & (w_1,\cdots, w_n, \top,\top,\cdots)
\end{array}\]
These functions are measurable and permit to define a sequence of restricted measure spaces which is indeed a filtration:
\[\begin{array}{lcl}
\natfilt & :: & \dN \rightarrow (\dB\ \stream)\ \measure\\
\natfilt\ n & = & \calM_{(\projt_n)}
\end{array}\]
We can thus define a locale for the infinite coin toss space along with this filtration:
\[\begin{array}{l}
\keyw{locale}\ \infiltr = \infct\ + \\
\keyw{fixes}\ \calF\
\keyw{assumes}\ \calF = \natfilt
\end{array}\]
Any other considered filtration on this probability space will be a sub-filtration of the natural filtration.
\hide{\Mnacho{todo: discuss remark ``Si j'ai bien compris, la filtration définie ici est celle dont l'existence est supposée précédemment page 7 ? C'est ce qui assure que les prix des derivatives ne soient pas définis de manière arbitraire ? A mon avis il faudrait expliquer cela un peu + en détail.''}}

\newcommand{\probcomp}{\nu}

An important feature of the natural filtration is that the expectation of any $\natfilt_n$-measurable function is very similar to that of a function on a finite probability space: for $\omega = \omega_1,\cdots, \omega_n,\cdots$ and $i\in \dN$, if $\probcomp_i(\omega) \isdef \keyw{if}\ \omega_i\ \keyw{then}\ p\ \keyw{else}\ 1-p$, then we have
\[\begin{array}{l}
\keyw{lemma}\ \textsc{expect-prob-comp} \\
\keyw{assumes}\ f\in \borelmeasurable\ \natfilt_n\\
\keyw{shows}\ \expect{f} = \sum_{\omega \in \projt_n(\Omega_\calM)}\left(\prod_{i = 1}^n \probcomp_i(\omega_i).f(\omega)\right)
\end{array}\]

\newcommand{\geomrw}{\texttt{geom-rand-walk}}
\newcommand{\crrmarket}{\texttt{CRR-market}}
\newcommand{\sproj}[1]{\lfloor#1\rfloor}
\newcommand{\procval}[2]{#1^{#2}}

In the CRR model, the price of the risky asset is modeled by a \emph{geometric random walk} with parameters specifying the upward and downward movements as well as the price of the asset at time $0$:
\[\begin{array}{lcl}
\geomrw & :: & \dR\rightarrow \dR\rightarrow \dR\rightarrow \\
& & \quad(\dN\rightarrow (\dB\ \stream)\rightarrow \dR)\\
(\geomrw\ u\ d\ v)\ 0\ \omega & = & v\\
(\geomrw\ u\ d\ v)\ (n+1)\ \omega & = & (\keyw{if}\ \omega_n\ \keyw{then}\ u\ \keyw{else}\ d)\ \times\\
&&\quad ((\geomrw\ u\ d\ v)\ n\ \omega)
\end{array}\]

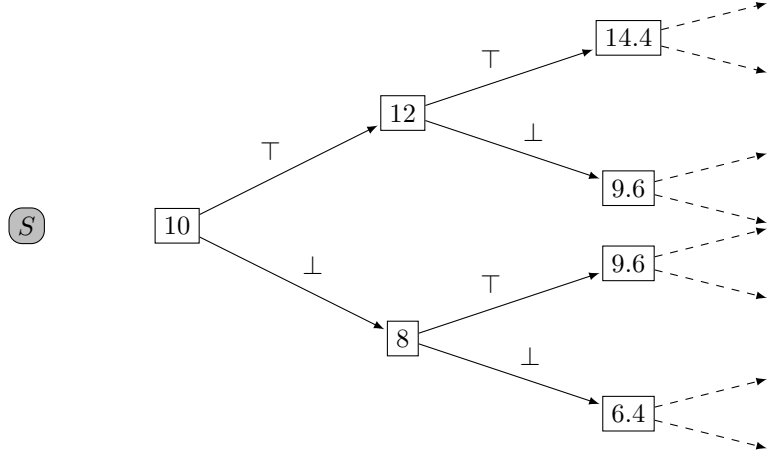
\begin{figure}[t]
	\begin{center}
		\begin{tikzpicture}[highlight/.style={draw=red, text=blue},shorten >=1pt,node distance=2cm,auto,>=stealth']
		\node[draw, rectangle,rounded corners = 5pt, fill=lightgray] (S)at(0,2.5){$S$};
		\node[draw, rectangle] (S0)at(2,2.5){$10$};
		\node[draw, rectangle] (Su)at(5,4){$12$};
		\node[draw, rectangle] (Suu)at(8,5){$14.4$};
		\node[draw, rectangle] (Sud)at(8,3){$9.6$};
		\node[draw, rectangle] (Sd)at(5,1){$8$};
		\node[draw, rectangle] (Sdu)at(8,2){$9.6$};
		\node[draw, rectangle] (Sdd)at(8,0){$6.4$};
		\node[draw=none, rectangle] (Nuuu)at(10,5.5){};
		\node[draw=none, rectangle] (Nuud)at(10,4.5){};
		\node[draw=none, rectangle] (Nudu)at(10,3.5){};
		\node[draw=none, rectangle] (Nudd)at(10,2.5){};
		\node[draw=none, rectangle] (NR)at(10,6){};
		\node[draw=none, rectangle] (Nduu)at(10,2.5){};
		\node[draw=none, rectangle] (Ndud)at(10,1.5){};
		\node[draw=none, rectangle] (Nddu)at(10,0.5){};
		\node[draw=none, rectangle] (Nddd)at(10,-0.5){};
		\node[draw=none] (Z)at(10,0) {};
		\draw[->,>=latex] (S0) -- (Su) node[midway] {{$\top$}};
		\draw[->,>=latex] (S0) -- (Sd) node[midway] {{$\bot$}};
		\draw[->,>=latex] (Su) -- (Suu) node[midway] {{$\top$}};
		\draw[->,>=latex] (Su) -- (Sud) node[midway] {{$\bot$}};
		\draw[->,>=latex] (Sd) -- (Sdu) node[midway] {{$\top$}};
		\draw[->,>=latex] (Sd) -- (Sdd) node[midway] {{$\bot$}};
		\draw[dashed, ->,>=latex] (Suu) -- (Nuuu);
		\draw[dashed, ->,>=latex] (Suu) -- (Nuud);
		\draw[dashed, ->,>=latex] (Sud) -- (Nudu);
		\draw[dashed, ->,>=latex] (Sud) -- (Nudd);
		\draw[dashed, ->,>=latex] (Sdu) -- (Nduu);
		\draw[dashed, ->,>=latex] (Sdu) -- (Ndud);
		\draw[dashed, ->,>=latex] (Sdd) -- (Nddu);
		\draw[dashed, ->,>=latex] (Sdd) -- (Nddd);
		\end{tikzpicture}
	\end{center}
\caption{Example of a geometric  random walk}\label{fig:grw}
\end{figure}

\begin{example}
	Figure \ref{fig:grw} depicts the first values of the geometric random walk process $\geomrw\ 1.2\ 0.8\ 10$. This is a process with a deterministic initial value 10. Intuitively, at time 1, if the outcome of a coin toss is a head, then this process has a value of 12, and if the outcome is a tail, then it has a value of 8. If at time 2 the first two outcomes are a head then a tail, then the value is 9.6, etc.
\end{example}

The geometric random walk process is an adapted process, in the infinite coin toss space equipped with its natural filtration, 
{since its value at time $n$ depends only on the outcome of the first $n$ coin tosses:}
\[\begin{array}{l}
	\keyw{lemma}\ \textsc{geom-rand-walk-borel-adapted}:\\
	\quad\quad \boreladapt\ (\geomrw\ u\ d\ v)
\end{array}\]

\newcommand{\probgrw}{\texttt{prob-grw}}
\newcommand{\geomproc}{\texttt{geom-proc}}

\[\begin{array}{l}
\keyw{locale}\ \probgrw = \infct\ + \\
\keyw{fixes}\ \geomproc\ \keyw{and}\ u\ \keyw{and}\ d\ \keyw{and}\ v\\
\keyw{assumes}\ \geomproc = \geomrw\ u\ d\ v
\end{array}\]

We define a locale in which there is a stochastic process that is a geometric random walk:

\newcommand{\crrhyps}{\texttt{CRR-hyps}}

The locale for the market in the CRR model is defined as follows:
\[\begin{array}{l}
\keyw{locale}\ \crrhyps\ =\ \probgrw + \rfreeasset +\\
\keyw{fixes}\ S\\
\keyw{assumes}\ \stocks\ \mkt = \set{S, \riskfreeasset}\\
\keyw{and}\ \prices\ \mkt\ S = \geomproc\\
\keyw{and}\ 0 < v\ \keyw{and}\ 0<d<u\\
\keyw{and}\ 0 < p < 1
\end{array}\]
In particular, we require that $0<p<1$ so that $S$ is indeed a risky asset.

\newcommand{\calG}{\mathcal{G}}

The filtration associated with this probability space is meant to represent the fact that the information available at time $n$ is the price evolution of the risky asset up to time $n$. 
We thus define a function that associates a filtration to a stochastic process $X$, such that at time $n$, the corresponding measure space is the smallest subalgebra for which $X_k$ is measurable for all $k\leq n$.
\newcommand{\stochprocfilt}{\texttt{stoch-proc-filt}}

\[\begin{array}{lcl}
\stochprocfilt & :: & \alpha\ \measure \rightarrow (\dN\rightarrow\alpha \rightarrow \beta) \rightarrow \beta\ \measure \rightarrow\\
& & \quad \dN \rightarrow \alpha\ \measure \\
\stochprocfilt\ M\ X\ N\ n & = & \isasigma\ \Omega_\calM\ \bigcup_{k\leq n} \setof{X_i^{-1}(A) \cap \Omega_\calM}{A\in \calA_\calN}\\
\end{array}\]

In the locale below, we denote by $\calG$ the filtration such that at time $n$, $\calG_n$ is the smallest subalgebra for which $\prices\ \mkt\ S\ k$ is borel-measurable for all $k$.

\[\begin{array}{l}
\keyw{locale}\ \crrmarket\ =\ \crrhyps +\\
\keyw{fixes}\ \calG\\
\keyw{assumes}\ \calG\ =\ \stochprocfilt\ \calM\ \geomproc\ \borel
\end{array}\]

In order to compute fair prices, the CRR market is required to be viable. We have the following result:

\[\begin{array}{l}
\keyw{lemma}\ \textsc{viable-iff} \\
\keyw{shows}\ \viable\ \mkt\ \Leftrightarrow\ (d < 1+r < u)
\end{array}\]

The direct implication is straightforward to prove. If for example the risky asset always has a return greater than the risk-free rate, i.e., $1+r \leq d$, then an arbitrage can be obtained by borrowing the initial value of the risky asset, $v$, at time 0 and buying one share of the risky asset. This results in a portfolio with initial value 0. At time $1$, the {\updval} of the portfolio is either $dv - (1+r)v$ or $uv - (1+r)v$; in both cases this value is positive, and it is strictly positive with probability $p>0$. The market can therefore not be viable. The proof of the other direction is not as obvious. Intuitively, this result can be proved by showing that when $d < 1+r < u$, if there is a scenario in which a portfolio with an initial value of 0 admits a strictly positive {\updval}, then there necessarily exists a scenario in which this portfolio admits a strictly negative value.

\label{viable_iff}

\newcommand{\crrviable}{\texttt{CRR-market-viable}}

We may thus define a locale for a viable CRR market:

\[\begin{array}{l}
\keyw{locale}\ \crrviable\ =\ \crrmarket +\\
\keyw{assumes}\ \viable\ \mkt
\end{array}\]

Next, we provide a necessary and sufficient condition for the existence of a risk-neutral bernoulli stream space that is filtration-equivalent to $\calM$.

\[\begin{array}{l}
\keyw{lemma}\ \textsc{risk-neutral-iff} \\
\keyw{assumes}\ \calN = \bernstream\ q\\
\keyw{and}\ 0 < q < 1\\
\keyw{shows}\ \riskneutralprob\ \calG\ \mkt\ r\ \calN\ \Leftrightarrow\ q = \frac{1+r-d}{u-d}
\end{array}\]

\hide{\Mnacho{I dont' know how to prove that $\calN$ is necessarily a bernoulli space, I didn't find any info on this topic}}

We also prove that every derivative is attainable in the CRR model:

\[\begin{array}{l}
\keyw{lemma}\ \textsc{CRR-market-complete}:\\
\keyw{shows}\ \complete
\end{array}\]

The result is proven by constructing a replicating portfolio for any $\calG_T$-measurable payoff $\pyf: \alpha \rightarrow \dR$ 
and exercise time $T$. 
Note that the fact that function $\pyf$ is 
$\calG_T$-measurable ensures that the payoff only depends on information available up to time $T$.
The principle of the construction of the portfolio is explained in details  on an example in Section \ref{sect:comp_ex}.

We finally obtain the final result:

\[\begin{array}{l}
\keyw{lemma}\ \textsc{CRR-market-fair-price}:\\
\keyw{assumes}\ \kappa\in \borelmeasurable\ \calG_T\\
\keyw{and}\ \calN = \bernstream\ \frac{1+r-d}{u-d}\\
\keyw{shows}\ \fairprice\ \mkt\\
\quad\quad \sum_{\omega \in \projt_T(\Omega_\calM)}\left(\prod_{i = 1}^T \probcomp_i(\omega_i).(\discval\ r\ \kappa\ \omega)\right) \\
\quad\quad \kappa\ T
\end{array}\]

\section{A complete example}


\label{sect:comp_ex}

We use the results above to price a \emph{lookback option} and illustrate how the completeness of the Cox-Ross-Rubinstein market is proved by constructing a replicating portfolio for this option. A lookback option is characterized by a  maturity $T$ and at this maturity, pays $\max_{0\leq i\leq T} S_i -S_T$. In other words, instead of having a payoff that only depends on the value of the risky asset at maturity, a lookback option has a payoff that depends on \emph{all} the values of the risky asset until maturity. It is called a \emph{path-dependent} option.

We assume that the risky asset has an initial value 10, an upward movement $u = 1.2$ and a downward movement $d = 0.8$. The risk-free rate is $r = 3\%$ (see Figure \ref{fig:lookback}). Consider a lookback option with maturity $T = 2$. Its payoff is depicted on the right-hand side of the figure. If the outcomes of the first two coin tosses are heads, then the maximal value of the risky asset is its value at time 2, so that this option does not pay anything. If the outcomes are first a head then a tail, then the value of the risky asset at time 2 is 9.6\euro{} the option pays off $2.4$\euro{}. Note that if the first two coin tosses are a tail then a head, then the value of the risky asset at time 2 is also $9.6$\euro{}, but the option only pays off $0.4$\euro{}.

The fair price of this option is computed using Lemma \textsc{CRR-market-fair-price}, which states that this fair price is the risk-neutral expectation of the discounted payoff of the option. The risk-neutral probability space is given by taking the probability of the coin landing on its head equal to $\frac{1+r - d}{u-d} = 0.575$. We thus obtain the following table:
\[\begin{array}{|c|c|c|c|c|c|}
\hline		
\textrm{Outcomes}  & \top\top & \top\bot & \bot\top & \bot\bot\\
\textrm{Probability} & 0.330625 & 0.244375 & 0.244375 & 0.180625\\
\hline
\text{Payoff}  & 0 & 2.4 & 0.4 & 3.6  \\	
\text{Disc. payoff}  & 0 & 2.262 & 0.377 & 3.393  \\
\hline
\end{array}\]
We deduce that the fair price of this option is $1.2579$\euro{}.

We now construct a replicating portfolio for this option. This portfolio will be constructed by going backward in time. First assume the outcome of the first coin toss is a head. In this scenario, we construct a portfolio that starts at time 1. The fair price of the option is given using the table below:
\[\begin{array}{|c|c|c|c|}
\hline		
\textrm{Outcomes}  & \top\top & \top\bot\\
\textrm{Probability} & 0.575 & 0.425  \\
\hline
\text{Payoff}  & 0 & 2.4  \\	
\text{Disc. payoff}  & 0 & 2.33  \\
\hline
\end{array}\]
We deduce that the fair price of the option (and the initial value of the portfolio under construction starting at time 1) is approximately 0.9903\euro{}. The quantity invested in the risky asset is given by 
\[\Delta_\top \isdef \frac{\kappa_{\top\top} - \kappa_{\top\bot}}{S_{\top\top} - S_{\top\bot}} = \frac{0 - 2.4}{14.4 - 9.6} = -0.5,\]
where $\kappa_{\omega_1\omega_2}$ and $S_{\omega_1\omega_2}$ respectively denote the payoff of the derivative and the value of the risky asset at time $2$, depending on the outcomes of the first two coin tosses $\omega_1$ and $\omega_2$.

This means that half a share of the risky asset is sold (a short sell) for 6\euro{}. Since the initial value of the portfolio is 0.9903\euro{}, this cash, along with the one obtained by selling the risky asset for a total of $6.9903$\euro{}, is invested in the risk-free rate. At time $2$, the cash invested in the risk-free rate is recovered and worth $6.9903 * 1.03 = 7.2$\euro{}; the half-share of the risky asset is bought back.
\begin{itemize}
	\item If the outcome of the second coin toss is a head, then the risky asset is worth 14.4\euro{}, so 7.2\euro{} are necessary to buy half the share back. The value of the portfolio is 0\euro{}.
	\item If the outcome of the second coin toss is a tail, then the risky asset is worth 9.6\euro{}, so 4.8\euro{} are necessary to buy half the share back. The value of the portfolio is 2.4\euro{}.
\end{itemize}

Now assume the outcome of the first coin toss is a tail. The fair price of the option at time 1 in this scenario is given using the table below:
\[\begin{array}{|c|c|c|c|}
\hline		
\textrm{Outcomes}  & \bot\top & \bot\bot\\
\textrm{Probability} & 0.575 & 0.425  \\
\hline
\text{Payoff}  & 0.4 & 3.6  \\	
\text{Disc. payoff}  & 0.38835 & 3.49515  \\
\hline
\end{array}\]
The fair price of the option under this scenario is approximately 1.7087\euro{}. The quantity invested in the risky asset is 
\[\Delta_\bot \isdef \frac{\kappa_{\bot\top} - \kappa_{\bot\bot}}{S_{\bot\top} - S_{\bot\bot}} = \frac{0.4 - 3.6}{9.6 - 6.4} = -1.\]
One share of the risky asset is sold for 8\euro{}, and the proceeds of this sale, along with the initial value of the portfolio are invested in the risk-free asset. At time $2$, the amount thus invested is worth 10\euro{}, and the share of the risky asset is bought back.
\begin{itemize}
	\item If the outcome of the second coin toss is a head, then the risky asset is worth 9.6\euro{}, so the value of the portfolio is 0.4\euro{}.
	\item If the outcome of the second coin toss is a tail, then the risky asset is worth 6.4\euro{}, so the value of the portfolio is 3.6\euro{}.
\end{itemize}

\begin{figure}[t]
	\begin{center}
		\begin{tikzpicture}[highlight/.style={draw=red, text=blue},shorten >=1pt,node distance=2cm,auto,>=stealth']
		\node[draw, rectangle,rounded corners = 5pt, fill=lightgray] (R)at(0,6){$R$};
		\node[draw, rectangle, fill = Dandelion] (R0)at(2,6){$1$};
		\node[draw, rectangle, fill = Dandelion] (R1)at(5,6){$1.03$};
		\node[draw, rectangle, fill = Dandelion] (R2)at(8,6){$1.061$};
		\node[draw, rectangle,rounded corners = 5pt, fill=lightgray] (S)at(0,2.5){$S$};
		\node[draw, rectangle, fill = Orchid] (S0)at(2,2.5){$10$};
		\node[draw, rectangle, fill = Orchid] (Su)at(5,4){$12$};
		\node[draw, rectangle, fill = Orchid] (Suu)at(8,5){$14.4$};
		\node[draw, rectangle, fill = Orchid] (Sud)at(8,3){$9.6$};
		\node[draw, rectangle, fill = Orchid] (Sd)at(5,1){$8$};
		\node[draw, rectangle, fill = Orchid] (Sdu)at(8,2){$9.6$};
		\node[draw, rectangle, fill = Orchid] (Sdd)at(8,0){$6.4$};
		\node[draw=none, rectangle] (Nuuu)at(10,5.5){};
		\node[draw=none, rectangle] (Nuud)at(10,4.5){};
		\node[draw=none, rectangle] (Nudu)at(10,3.5){};
		\node[draw=none, rectangle] (Nudd)at(10,2.5){};
		\node[draw=none, rectangle] (NR)at(10,6){};
		\node[draw=none, rectangle] (Nduu)at(10,2.5){};
		\node[draw=none, rectangle] (Ndud)at(10,1.5){};
		\node[draw=none, rectangle] (Nddu)at(10,0.5){};
		\node[draw=none, rectangle] (Nddd)at(10,-0.5){};
		\node[draw, rectangle, fill = SeaGreen] (Puu)at(10,5){$0$};
		\node[draw, rectangle, fill = SeaGreen] (Pud)at(10,3){$2.4$};
		\node[draw, rectangle, fill = SeaGreen] (Pdu)at(10,2){$0.4$};
		\node[draw, rectangle, fill = SeaGreen] (Pdd)at(10,0){$3.6$};
		\node[draw=none] (Z)at(10,0) {};
		\draw[->,>=latex] (R0) -- (R1);
		\draw[->,>=latex] (R1) -- (R2);
		\draw[->,>=latex] (S0) -- (Su) node[midway] {{$\top$}};
		\draw[->,>=latex] (S0) -- (Sd) node[midway] {{$\bot$}};
		\draw[->,>=latex] (Su) -- (Suu) node[midway] {{$\top$}};
		\draw[->,>=latex] (Su) -- (Sud) node[midway] {{$\bot$}};
		\draw[->,>=latex] (Sd) -- (Sdu) node[midway] {{$\top$}};
		\draw[->,>=latex] (Sd) -- (Sdd) node[midway] {{$\bot$}};
		\draw[dashed, ->,>=latex] (R2) -- (NR);
		\draw[dashed, ->,>=latex] (Suu) -- (Nuuu);
		\draw[dashed, ->,>=latex] (Suu) -- (Nuud);
		\draw[dashed, ->,>=latex] (Sud) -- (Nudu);
		\draw[dashed, ->,>=latex] (Sud) -- (Nudd);
		\draw[dashed, ->,>=latex] (Sdu) -- (Nduu);
		\draw[dashed, ->,>=latex] (Sdu) -- (Ndud);
		\draw[dashed, ->,>=latex] (Sdd) -- (Nddu);
		\draw[dashed, ->,>=latex] (Sdd) -- (Nddd);
		\end{tikzpicture}
	\end{center}
	\caption{Lookback option settings and payoff. $R$ denotes the risk-free asset and $S$ the risky one.}\label{fig:lookback}
\end{figure}
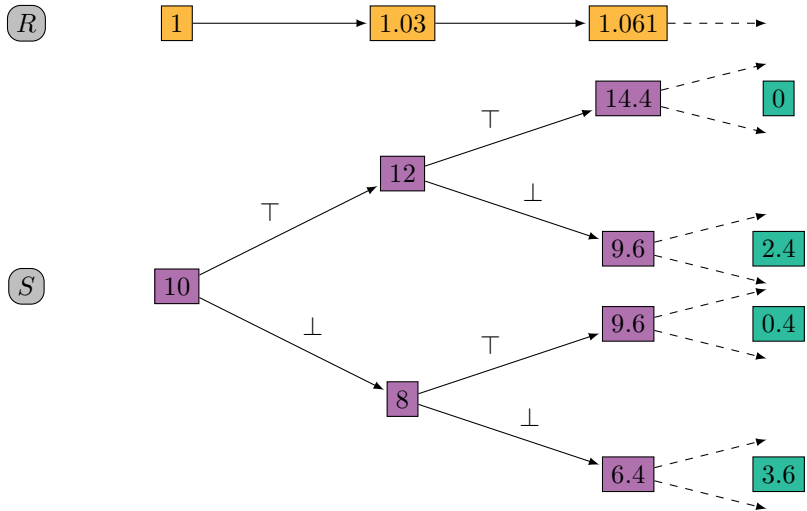

We now construct a portfolio with initial value $1.2579$\euro{}, and worth $0.9903$\euro{} if the outcome of the first coin toss is a head, and $1.7087$\euro{} if the outcome is a tail. The quantity invested in the risky asset is given by 
\[\Delta_\top \isdef \frac{0.9903 - 1.7087}{12-8} = -0.1796.\] 
This quantity of the risky asset is sold for 10\euro{}, and the proceeds are invested in the risk-free asset, along with the initial value of the portfolio. At time 1, the cash thus invested is worth $3.145517$\euro{}. The quantity of risky asset that was shorted is bought back.
\begin{itemize}
	\item If the outcome of the first coin toss is a head, then the risky asset is worth 12\euro{} and buying back the quantity that was shorted costs $2.1552$\euro{}, so the value of the portfolio is $0.9903$\euro{}.
	\item If the outcome of the first coin toss is a tail, then the risky asset is worth 8\euro{} and buying back the quantity that was shorted costs $1.4368$\euro{}, so the value of the portfolio is $1.7087$\euro{}.
\end{itemize}

To recap, the seller of the lookback option sells it for $1.2579$\euro{}, and constructs a replicating portfolio as follows.
\begin{enumerate}
	\item The seller receives $1.796$\euro{} by short selling the risky asset and invests the $3.0539$\euro{} in the risk-free asset until time 1.
	\item If at time 1 the outcome of the first coin toss is a head, then the seller uses the {\updval} of the portfolio, 0.9903\euro{}, to short half a share of the risky asset and invest $6.9903$\euro{} in the risk-free asset. Otherwise, the seller uses the {\updval} of the portfolio, $1.7087$\euro{}, to short one share of the risky asset and invest $9.7087$\euro{} in the risk-free asset.
	\item At time 2, quantity of risky asset that was shorted is bought back and the cash invested in the risk-free asset is withdrawn; the {\updval} of the portfolio is exactly equal to the payoff of the lookback option.
\end{enumerate}

	This construction can be generalized to arbitrary $\natfilt_T$-measurable functions. 
At any time $t < T$, the composition of the portfolio is determined in such a way that
its {\updval} at time $t+1$ matches the value already computed at time $t+1$, for both outcomes 
of the next coin toss.
This yields a system of two linear equations, one for each possible outcome of the toss coin, 
with two variables (the amounts of risk-free and risky assets, respectively). 
Lemma \textsc{viable-iff} on Page \pageref{viable_iff} imposes additional conditions on $u,d,r$ that ensure that 
the system  admits a unique solution.

\section{Discussion}

We have formalized a framework for proving financial results in Isabelle. The formalization permits a formal definition of fair prices in Isabelle, and presents one of the main pricing results in finance: under a risk-neutral probability, the fair price of an attainable derivative is equal to the expectation of its discounted payoff. This formalization is quite extensive, as many financial notions had to be introduced, and it was used to prove that every derivative admits a fair price in the Cox-Ross-Rubinstein model of an equity market, by proving the completeness of this market. As far as future work is concerned, we intend to work on the pricing in the Cox-Ross-Rubinstein model of American options, that can be exercised at any time by the buyer until the maturity --and not simply at maturity, as for European options--. Pricing such options will require the definition of additional notions, such as \emph{supermartingales}, and our aim will be to implement a completely certified pricer for such options. We also intend to pursue our formalization effort of mathematical finance and extend our results to a continuous-time setting. This is an ambitious and interesting task, and we hope this first formalization will encourage other researchers, from computer science and financial mathematics to extend these results in Isabelle.


\bibliography{biblio}

\begin{thebibliography}{10}

\bibitem{CentralLimit}
J.~Avigad, J.~H{\"{o}}lzl, and L.~Serafin.
\newblock A formally verified proof of the central limit theorem.
\newblock {\em J. Autom. Reasoning}, 59(4):389--423, 2017.

\bibitem{BlackScholes}
F.~Black and M.~Scholes.
\newblock The pricing of options and corporate liabilities.
\newblock {\em Journal of political economy}, 81(3):637--654, 1973.

\bibitem{Blanchette13}
J.~C. Blanchette.
\newblock Relational analysis of (co)inductive predicates, (co)algebraic
  datatypes, and (co)recursive functions.
\newblock {\em Software Quality Journal}, 21(1):101--126, 2013.

\bibitem{CRR}
J.~C. Cox, S.~A. Ross, and M.~Rubinstein.
\newblock Option pricing: A simplified approach.
\newblock {\em Journal of Financial Economics}, 7(3):229--263, 1979.

\bibitem{Durrett}
R.~Durrett.
\newblock {\em Probability : theory and examples}.
\newblock The Wadsworth \& Brooks/Cole statistics/probability series. Wadsworth
  Inc. Duxbury Press, Belmont, California, 1991.

\bibitem{EPCRR}
M.~Echenim and N.~Peltier.
\newblock The binomial pricing model in finance: {A} formalization in isabelle.
\newblock In L.~de~Moura, editor, {\em Automated Deduction - {CADE} 26 - 26th
  International Conference on Automated Deduction, Gothenburg, Sweden, August
  6-11, 2017, Proceedings}, volume 10395 of {\em Lecture Notes in Computer
  Science}, pages 546--562. Springer, 2017.

\bibitem{hoelzl2012thesis}
J.~H{\"o}lzl.
\newblock {\em Construction and Stochastic Applications of Measure Spaces in
  Higher-Order Logic}.
\newblock PhD thesis, Institut f{\"u}r Informatik, Technische Universit{\"a}t
  M{\"u}nchen, October 2012.

\bibitem{HolzlJAR}
J.~H{\"o}lzl.
\newblock Markov chains and markov decision processes in isabelle/hol.
\newblock {\em Journal of Automated Reasoning}, pages 1--43, 2016.
\newblock {\url{http://dx.doi.org/10.1007/s10817-016-9401-5}}.

\bibitem{HolzlLT15}
J.~H{\"{o}}lzl, A.~Lochbihler, and D.~Traytel.
\newblock A {F}ormalized {H}ierarchy of {P}robabilistic {S}ystem {T}ypes -
  {P}roof {P}earl.
\newblock In {\em Proc. of ITP}, pages 203--220, 2015.

\bibitem{Hull}
J.~Hull.
\newblock {\em Options, Futures and Other Derivatives}.
\newblock Pearson/Prentice Hall, 2009.

\bibitem{Merton}
R.~Merton.
\newblock The theory of rational option pricing.
\newblock {\em Bell Journal of Economics and Management Science}, 4:141--183,
  1973.

\bibitem{Nipkow:2002:IPA:1791547}
T.~Nipkow, M.~Wenzel, and L.~C. Paulson.
\newblock {\em Isabelle/HOL: A Proof Assistant for Higher-order Logic}.
\newblock Springer-Verlag, Berlin, Heidelberg, 2002.

\bibitem{Shreve}
S.~E. Shreve.
\newblock {\em Stochastic Calculus for Finance I: The Binomial Asset Pricing
  Model}.
\newblock Springer Finance, 2003.

\end{thebibliography}
\bibliographystyle{abbrv}
\end{document}